\documentclass[preprint,12pt]{elsarticle}

\usepackage{graphicx}

\usepackage[caption=false,font=footnotesize]{subfig}
\usepackage{amsmath}
\usepackage{amssymb}
\usepackage{amsthm}

\newtheorem{remark}{Remark}
\newtheorem{lemma}{Lemma}
\newtheorem{theorem}{Theorem}

\newtheorem{corollary}{Corollary}

\usepackage{url}
\usepackage{algorithm}
\usepackage{algorithmic}

\usepackage{IEEEtrantools}

\biboptions{sort}

\journal{Ad Hoc Networks}

\begin{document}

\begin{frontmatter}



\title{Secrecy Transmission Capacity in Noisy Wireless Ad Hoc Networks}


\author[FUN]{Jinxiao~Zhu \corref{cor1}}
\ead{jxzhu1986@gmail.com}
\author[FUN]{Yin~Chen }
\ead{ychen1986@gmail.com}
\author[XIDIAN]{Yulong~Shen}
\ead{ylshen@mail.xidian.edu.cn}
\author[FUN]{Osamu~Takahashi }
\ead{Osamu@fun.ac.jp}
\author[FUN]{Xiaohong~Jiang \corref{cor2}}
\ead{jiang@fun.ac.jp}
\author[TOHOUKU]{Norio~Shiratori}
\ead{norio@shiratori.riec.tohoku.ac.jp}


\cortext[cor1]{Corresponding author. Tel.: 081-0138-34-6226}
\cortext[cor2]{Principal corresponding author.}

\address[FUN]{School of Systems Information Science, Future University Hakodate, 041-8655 Japan.}
\address[XIDIAN]{State Key Laboratory of Integrated Services networks (ISN), Xidian University, 710071 P.R.China.}
\address[TOHOUKU]{The author is with GITS, Waseda University, Tokyo, 169-0051 Japan and RIEC, Tohoku University, Sendai-shi, 980-8579 Japan.}

\begin{abstract}
This paper considers the transmission of confidential messages over noisy wireless ad hoc networks, where both background noise and interference from concurrent transmitters affect the received signals.
For the random networks where the legitimate nodes and the eavesdroppers are distributed as Poisson point processes, we study the secrecy transmission capacity (STC), as well as the connection outage probability and secrecy outage probability, based on the physical layer security.
We first consider the basic fixed transmission distance model, and establish a theoretical model of the STC.
We then extend the above results to a more realistic random distance transmission model, namely nearest receiver transmission.
Finally, extensive simulation and numerical results are provided to validate the efficiency of our theoretical results and illustrate how the STC is affected by noise, connection and secrecy outage probabilities, transmitter and eavesdropper densities, and other system parameters.
Remarkably, our results reveal that a proper amount of noise is helpful to the secrecy transmission capacity.
\end{abstract}


\end{frontmatter}


\section{Introduction} \label{section1}

The inherent openness of wireless medium makes information security one of the most important and difficult problems in wireless networks.
Traditionally, information security is ensured by applying cryptography which encrypts a plain message into a ciphertext that is computationally infeasible for any adversary without the key to break (decrypt).
However,
due to the improvement in computing technology and complication in cryptographic key management,
there is an increasing concern that the cryptography no longer suffices, especially in sensitive applications requiring everlasting secrecy.
Recently, the physical layer security has been widely demonstrated as a promising approach to providing everlasting secrecy.
Unlike the traditional cryptography that ignores the difference between transmitting channels, the recent physical layer security achieves information-theoretic security by properly designing wiretap channel code according to the channel capacities \cite{wyner1975, Gaussian1978}
such that the original data can be hardly recovered by the eavesdropper regardless of how strong the eavesdropper's computing power is.

By now, a lot of research works have been dedicated to understand the performance of physical layer security.
Wyner initially studied the maximum secret information rate, namely secrecy capacity, for a discrete memoryless wire-tap channel,
where only three nodes are involved (one transmitter, one legitimate receiver and one eavesdropper),
and showed the existence of channel codes to ensure the message is reliably delivered to the legitimate receiver while secured at the eavesdropper \cite{wyner1975}.
Wyner's work was then extended to other channel models, such as Gaussian wire-tap channel \cite{Gaussian1978}, fading wire-tap channel with or without channel correlations \cite{Bloch2008, Jeon2011, Zhu2013, Zhu2014}, broadcast channels with confidential messages \cite{Csiszar1978}, etc.
Based on these pioneering works on the basic point-to-point wire-tap channels, many recent research efforts have been conducted to understand the performances of physical layer security on large-scale wireless networks, where lots of legitimate nodes and eavesdroppers are involved, in terms of secrecy throughput capacity \cite{Koyluoglu2012, Liang2011, Pinto2010a, Vasudevan2010}, secrecy coverage \cite{Sarkar2010}, connectivity \cite{Haenggi2008, Pinto2012, Goel2010,  Zhou2011a} and percolation phenomenon \cite{Pinto2010a, Sarkar2011, Sarkara2013} under secrecy constraints, etc.

This paper focuses on the study of secrecy transmission capacity (STC) in large-scale wireless networks, which is defined as the achievable rate of successful transmission of confidential messages per unit area of a network, subject to constraints on both connection outage probability and secrecy outage probability.
It is notable that the STC indicates the area spectral efficiency (ASE) of wireless networks under the given constraints on the levels of reliability and security, and hence it is of fundamental importance and can serve as a guideline for the design and development of wireless networks.
Besides, compared with the aforementioned studies on the secrecy throughput capacity of large-scale wireless networks that only provide scaling law results \cite{Koyluoglu2012, Liang2011, Pinto2010a, Vasudevan2010}, exact results can be obtained from STC study, which can lead to a finer optimization on network performance.

Some prior works on STC have been done by Zhou \textit{et al.} in \cite{Zhou2011b, Zhou2012}, where they calculated the secrecy transmission capacity for decentralized wireless networks with a fixed distance transmission scheme under the signal-to-interference ratio (SIR) model that neglects the impact of background noise.
It is noticed that the background noise is a ubiquitous natural phenomenon and ignoring it may cause inaccuracy in the performance estimation.
Moreover, it is also noticed that the additional noise on one hand is harmful to the reliability of a transmission since it makes the signal received at the intended receiver worse, on the other hand is helpful to the security performance since it makes the signal received at eavesdroppers worse. Hence, a natural question what is the overall impact of the noise on the STC.
Accordingly, a new dedicated study is still required to investigate the exact STC in wireless networks under the impact of background noise.

In this work, we focus on the secrecy transmission capacity in noisy wireless ad hoc networks where interference from concurrent transmitters and background noise from natural and sometimes man-made sources affect the received signals. The main contributions of this paper are as follows.
\begin{itemize}
  \item Based on the tools from stochastic geometry, we start the analysis from a basic fixed transmission distance scenario where each transmitter has an intended receiver at a fixed distance which is the same for all transmitters.
      We establish a general theoretical model of the STC, as well as the connection outage probability and secrecy outage probability, under the signal-to-interference-noise ratio (SINR) model.
      Furthermore, for the special scenario when the path-loss exponent $\alpha=4$ and noise power is the same across space and time slot, we derive a closed-form STC and then propose the condition to achieve a positive STC.
  \item We then extend the analysis of STC to a more realistic random transmission scenario, nearest receiver transmission in particular, and present the corresponding connection outage probability and STC.
      It is noticed that the transmit distance has no impact on the secrecy outage probability.
  \item Finally, we provide extensive simulation and numerical results to validate the efficiency of our theoretical models and also to illustrate our theoretical findings.
      Remarkably, our results indicate that a proper amount of noise can be helpful to the secrecy transmission capacity.
\end{itemize}

The remainder of this paper is organized as follows. Section~\ref{section2} presents the system model and performance metrics based on the physical layer security. In Section~\ref{section3}, we obtain analytical results on the secrecy transmission capacity for fixed transmission distance scenario.
Then Section~\ref{section4} extends the analysis to nearest receiver transmission scenario.
In Section~\ref{section5}, we validate the theoretical models by simulations and analyze the tradeoff between the system parameters.
Finally, concluding remarks are given in Section~\ref{section6}.

\section{System Model and Performance Metrics} \label{section2}
In this section, we introduce the basic system model of this paper and the performance metrics based on the physical layer security.
The notation and symbols used throughout the paper are summarized in Table \ref{table_notation}.

\subsection{System Model} \label{section_systemModel}
We consider an ad hoc wireless network consisting of both legitimate nodes and eavesdroppers over a two-dimensional Euclidean space $\mathbb{R}^2$.
For each time snapshot, locations of legitimate nodes are modeled as a homogeneous Poisson point process (PPP) $\Phi$ with density $\lambda$, denoted by $\Phi = \{X_i\}$, where $X_i\in\mathbb{R}^2$ is the location of the legitimate node $i$,
and locations of eavesdroppers are modeled as a PPP $\Phi_e$ with density $\lambda_e$, denoted by $\Phi_e = \{X_e\}$,  where $X_e\in\mathbb{R}^2$ is the location of the eavesdropper node $e$.
The PPP model for node locations is suitable when the nodes are independently and uniformly distributed over the network area, which is often reasonable for networks with indiscriminate node placement or substantial mobility \cite{Weber2010}.
The slotted ALOHA is employed at legitimate nodes as the medium access control (MAC) protocol. That is, in each time slot, each legitimate node independently decides to transmit with probability $p$ or act as a potential receiver otherwise. Hence, in each time slot, the set of all transmitters forms a PPP $\Phi_\textrm{T}$ with density $\lambda_\textrm{T}=p \lambda $ and the set of all receivers forms a PPP $\Phi_\textrm{R}$ with density $\lambda_\textrm{R}=(1-p) \lambda $. Notice that $\Phi = \Phi_\textrm{T} \cup \Phi_\textrm{R}$.

%
\begin{table}[!t]
\renewcommand{\arraystretch}{1.3}
\caption{Summary of Notations}
\label{table_notation}
\centering
\begin{tabular}{|l|l|}
\hline
Symbol & Meaning\\
\hline
$\Phi$ & Poisson point process (PPP) of legitimate node locations\\
$\Phi_e$ & PPP of eavesdropper locations\\
$\Phi_\textrm{T}$, $\Phi_\textrm{R}$ & Sets of transmitter and receiver locations, resp. \\
$\lambda$, $\lambda_e$ & Density of $\Phi$ and $\Phi_e$, resp.\\
$\lambda_\textrm{T}$, $\lambda_\textrm{R}$ & Density of $\Phi_\textrm{T}$ and $\Phi_\textrm{R}$ , resp. ($\lambda = \lambda_\textrm{T} + \lambda_\textrm{R}$)\\
$\textrm{P}_\textrm{co}$, $\textrm{P}_\textrm{so}$ & Connection and secrecy outage probability, resp.\\
$\sigma$, $\epsilon$ & Constraints on connection and secrecy outage probability,resp.\\
$\beta_t$, $\beta_e$ & SIR threshold for legitimate nodes and eavesdroppers, resp.\\
$\mathcal{R}_t$, $\mathcal{R}_s$ & Codewords rate and secrecy rate, resp.\\
$\mathcal{R}_e$ & Rate loss for securing the message against eavesdropping\\
$W$, $w$ & Random and fixed noise, resp.\\
$\alpha > 2$ & Path loss exponent\\
$H_{ij}$ & Power gain of the channel from node $i$ to node $j$ \\
$X_i$ & Location of node $i$\\
$|X_{i}|$ & Distance from node $i$ to the origin\\
$| X_{ij}|$ & Distance from node $i$ to node $j$\\
$\mathbb{P}(\cdot)$ & Probability operator \\
$\mathbb{E}(\cdot)$ & Expectation operator \\
\hline
\end{tabular}
\end{table}

In this paper, we assume that all transmitters use the same transmission power $\rho$, and that the intended receiver (and thus the transmit distance) for each transmitter will be determined independently in Section \ref{section3} and Section \ref{section4} for fixed and random transmit distances, respectively.
The signal propagation over the wireless medium is assumed to be affected by the large-scale path loss, the small-scale fading and an additional noise. The large-scale path loss is assumed to be $r^{-\alpha}$ over distance $r$, where $\alpha > 2$ is the path loss exponent\footnote{Usually, the path loss exponent is in the range of $3-5$ \cite{Rappaport1996}.}.
For the small-scale fading, we assume the channel follows the Rayleigh fading with unit mean and the fading coefficient is independent from path to path.
Hence, the signal power received at a receiver $j$ from a transmitter $i$ is given by $\rho H_{ij} |X_{ij}|^{-\alpha}$, where $H_{ij}$ and $|X_{ij}|$ are the channel fading gain and the distance between the nodes $i$ and $j$, respectively.
The additional noise power\footnote{The noise is a summation of unwanted or disturbing energy from natural and sometimes man-made sources, like industrial and aircraft noises.} at a given location for a given time slot is denoted by the random variable $W$, which is independent of $\Phi$.
The detection performance is characterized by the signal-to-interference-noise ratio (SINR), i.e., the ratio of signal power over interference plus noise power.

\subsection{Physical Layer Security and Performance Metrics} \label{section2_metric}
In the considered network, a transmitter wants to send confidential messages to its receiver in the hope that the messages are reliably received by the the receiver while secured against eavesdroppers.
For the secure encoding schemes, we consider the physical layer security that implemented by the well-known Wyner code \cite{wyner1975}.
Specifically, the Wyner's encoding scheme requires a transmitter to choose two rates, namely, the codeword rate $\mathcal{R}_t$ and secrecy rate $\mathcal{R}_s$.
It is noticed that $\mathcal{R}_s \leq \mathcal{R}_t$, and the rate difference between the two rates, denoted by $\mathcal{R}_e = \mathcal{R}_t -\mathcal{R}_s$, indicates the rate cost of securing message transmissions against eavesdropping.
For any transmitted message, the receiver is able to decode it with an arbitrarily small error probability if $\mathcal{R}_t$ is \textit{less} than the capacity of the channel from the transmitter to this receiver, while an eavesdropper is \textit{not} expected to recover it correctly if $\mathcal{R}_e$ is \textit{larger} than the capacity of the channel from the transmitter to this eavesdropper.
In this work, we focus on the scenario that all transmitters choose the same pair of  $\mathcal{R}_t$ and $\mathcal{R}_s$ (and thus $\mathcal{R}_e$), which is reasonable since the network is homogeneous.
For more details about the Wyner's encoding scheme, please refer to \cite{Zhu2014, Zhou2011}.

Based on the above physical layer security method, the following three performance metrics are studied in this paper:
\begin{itemize}
  \item \textbf{Connection outage probability (COP)}: We call connection outage happens when the SINR at the intended receiver is below a given threshold $\beta_t$. The connection outage probability, denoted by $\textrm{P}_\textrm{co}(\beta_t)$, is then defined as the probability that connection outage happens.
It is noticed that $\mathcal{R}_t$ is related with $\beta_t$ by $\beta_t = 2^{\mathcal{R}_t}-1$.
  \item \textbf{Secrecy outage probability (SOP)}: We call secrecy outage happens when the SINR at one or more eavesdroppers is above a given threshold $\beta_e$. The secrecy outage probability, denoted by $\textrm{P}_\textrm{so}(\beta_e)$, is then defined as the probability that secrecy outage happens. It is noticed that $\mathcal{R}_e$ is related with $\beta_e$ by $\beta_e = 2^{\mathcal{R}_e}-1$.
  \item \textbf{Secrecy transmission capacity (STC)}: It is defined as the achievable rate of successful transmission of confidential messages per unit area, for a given connection outage probability $\textrm{P}_\textrm{co}(\beta_t)=\sigma$ and a given secrecy outage probability $\textrm{P}_\textrm{so}(\beta_e)=\epsilon$:
\begin{equation}\label{eq_secTransCapacity_formula}
    \tau = (1-\sigma) \lambda_\textrm{T} \mathcal{R}_s.
\end{equation}
Notice that the secrecy rate $\mathcal{R}_s = \left[ \mathcal{R}_t - \mathcal{R}_e \right]^{+}$ is a function of both $\sigma$ and $\epsilon$.
\end{itemize}
It is notable that the connection outage probability gives a measure of the reliability level while the secrecy outage probability gives a measure of the security level.
The secrecy transmission capacity, which was first defined in \cite{Zhou2011b}, is a measure of spatial intensity of successful transmission rate of confidential messages under a reliability constraint and a secrecy constraint.

\section{First Model: Fixed Transmission Distance} \label{section3}
In this section, we present the COP, SOP and STC under the basic fixed transmission distance assumption, i.e., each transmitter is assumed to have a prearranged intended receiver at a fixed distance $L$ away.
This assumption has been widely adopted in the literary of transmission capacity \cite{Weber2005, Weber2010, Ganti2011}.
The extension to random distance will be given in Section \ref{section4}.

To evaluate the COP, we will condition on a typical transmitter at the origin $o$.
The distribution of the point process $\Phi_\textrm{T}$ is unaffected by the addition of a transmitter node at the origin by Slivnyak's Theorem \cite{Stoyan1996}.
Given this typical transmitter, we shift the origin $o$ to its intended receiver at $L$ away, which is called the typical receiver, and analyze the SINR at the receiver.
This conditional distribution is sometimes referred to as the Palm distribution, and since the network is homogeneous, the interference (and thus SINR) measured at the origin is a representative of the interference (and thus SINR) seen by all other receiver nodes in the network.

The SINR at the typical receiver located at $o$ is given by
\begin{equation}\label{SINR_origin}
    \textrm{SINR}_0 = \frac{\rho H_{0} L^{-\alpha}}{W + I_0},
\end{equation}
where $I_0 = \Sigma_{k \in \Phi_\textrm{T} } \rho  H_{k0} |X_{k}| ^{-\alpha}$ is the interference at the origin, $H_{0}$ is the channel fading gain between the typical transmitter and receiver, $H_{k0}$ and $| X_{k}|$ are the channel fading gain and the distance between the interferer at $X_{k}$ and the typical receiver at the origin, respectively.

Based on the definition in Section \ref{section2_metric}, the COP can be derived by
\begin{equation}\label{conOutProDef}
    \textrm{P}_\textrm{co}(\beta_t) = \mathbb{P}(\textrm{SINR}_0 < \beta_t) = 1- \mathbb{P}(\textrm{SINR}_0 \geq \beta_t).
\end{equation}
For the exponential $H_0$ and random noise $W$, the success probability of transmission in an infinite planar network without eavesdroppers has been derived in \cite{Baccelli2006}. Following the similar method as that of deriving the success probability, the COP can be given by \begin{equation} \label{eq_conOutPro}
    \textrm{P}_\textrm{co} (\beta_t) = 1- \exp \left[ - \theta \left(\frac{\beta_t}{\rho}\right)^{\frac{2}{\alpha}} L^2 \right] \mathcal{L}_{W} \left( \frac{\beta_t}{\rho} L^\alpha \right),
\end{equation}
where  $\theta = \pi \lambda_\textrm{T} \Gamma( 1- 2/\alpha ) \Gamma( 1+ 2/\alpha )$, $\Gamma(\cdot)$ is the Gamma function and $\mathcal{L}_{W}(\cdot)$ is the Laplace transform of $W$.

We now shift the origin back to the typical transmitter node and consider the SOP.
Consider a transmission from the typical transmitter to an eavesdropper $e$, the received SINR at $e$ is given by
\begin{equation}\label{SINR_e}
    \textrm{SINR}_e = \frac{ \rho H_{e}| X_{e} | ^{-\alpha}}{\Sigma_{k \in \Phi_\textrm{T}} \rho H_{ke} | X_{ke} | ^{-\alpha} + W},
\end{equation}
where $H_{e}$ and $|X_{e}|$ are the channel fading gain and the distance between the typical transmitter and the eavesdropper $e$, respectively, $H_{ke}$ and $| X_{ke}|$ are the channel fading gain and the distance between the interferer $k$ and the eavesdropper $e$, respectively.

According to the definition in Section \ref{section2_metric}, secrecy outage happens if any one of eavesdroppers is able to recover the transmitted message. Let $E = \{ e \in \Phi_e : \textrm{SINR}_{e} > \beta_e \}$ be the set of eavesdroppers that can cause secrecy outage. Define an indicator function $1_{E}(e)$, which equals to $1$ if $e \in E$ and equals to $0$ otherwise. Then $\prod_{e \in \Phi_e}(1 - 1_{E}(e))$ is equal to 1 if the transmission from the typical transmitter is secured against any eavesdropper. Hence, the SOP can be obtained by

\begin{IEEEeqnarray}{rCl}\label{eq_secOutPro_formula}
    \textrm{P}_\textrm{so} (\beta_e) \! &=& \! 1 - \mathbb{E}_{\Phi_l} \left\{ \mathbb{E}_{\Phi_e} \!\! \left\{ \mathbb{E}_{H} \left\{ \prod_{e \in \Phi_e} \left( 1 - 1_{E}(e) \right ) \right\} \right\} \right\}  \nonumber \\
   \! &\stackrel{(a)}{=}& \! 1 \! - \! \mathbb{E}_{\Phi_l} \! \left\{ \mathbb{E}_{\Phi_e} \! \! \left\{ \prod_{e \in \Phi_e} \!\! \left( 1 - \mathbb{P} \left(\textrm{SINR}_{e} \! \geq \! \beta_e| \Phi_e,\Phi_l \right ) \right )  \right\} \right\}, \nonumber \\
\end{IEEEeqnarray}
where ($a$) is due to the assumption that the fading coefficient is independent from path to path.
Thus, the SOP can be derived in the following lemma.
\begin{lemma} \label{lemma_secOutPro}
For the concerned wireless network with network parameters $\lambda_\textrm{T}$, $\lambda_e$, $W$ and $\alpha$, and transmission parameter $\rho$ defined above,
its secrecy outage probability for a given eavesdroppers' SINR threshold $\beta_e$
is upper bounded by
\begin{equation} \label{eq_secOutPro_ub1}
    \textrm{P}_\textrm{so}^{u}(\beta_e) = 1 - \exp \! \left[ - 2 \pi \lambda_e \!\! \int_{0}^{\infty} \!\! e^{- \left( \frac{\beta_e}{\rho}\right )^{\frac{2}{\alpha}} \theta r^2 } \mathcal{L}_{W} \!\! \left(\frac{\beta_e}{\rho} r^\alpha \! \right) r \mathrm{d}{r} \right],
\end{equation}
and lower bounded by
\begin{equation} \label{eq_secOutPro_lb1}
    \textrm{P}_\textrm{so}^{l}(\beta_e) = 2 \pi \lambda_e \!\! \int_{0}^{\infty} \!\! e^{ - \left(  \theta \left( \frac{\beta_e}{\rho}\right )^{\frac{2}{\alpha}} + \pi \lambda_e\right )r^2  } \mathcal{L}_{W} \!\! \left(\frac{\beta_e}{\rho} r^\alpha \! \right) r \mathrm{d}{r},
\end{equation}
where $\theta = \pi \lambda_\textrm{T} \Gamma( 1- 2/\alpha ) \Gamma( 1+ 2/\alpha )$ is the same as defined above.
\end{lemma}
\begin{IEEEproof}
Based on the secrecy outage formula in (\ref{eq_secOutPro_formula}), we have
\begin{IEEEeqnarray}{rCl}\label{eq_secOutPro_ub_derivation}
    \textrm{P}_\textrm{so}(\beta_e)  &\stackrel{(b)}{=}& 1 - \mathbb{E}_{\Phi_l} \! \left\{ \exp \! \left[ - \lambda_e \!\! \int_{\mathbb{R}^2 } \! \mathbb{P} \left(\textrm{SINR}_e \geq \beta_e| \Phi_l \right ) \mathrm{d}{X_e} \right] \! \right\}  \nonumber \\
    &\stackrel{(c)}{\leq}& 1 - \exp \left[ - \lambda_e \int_{\mathbb{R}^2} \mathbb{P} \left(\textrm{SINR}_e \geq \beta_e \right ) \mathrm{d}{X_e} \right] \nonumber \\
    &\stackrel{(d)}{=}& 1 - \exp \! \left[ - 2 \pi \lambda_e     \vphantom{\left(\frac{\beta_t}{\rho}\right)^{\frac{2}{\alpha}}} \right. \nonumber \\
    && \left. \int_{0}^{\infty} \!\! \exp \left[- \theta \left(\frac{\beta_e}{\rho}\right)^{\frac{2}{\alpha}} r^2 \right] \mathcal{L}_{W} \left(\frac{\beta_e}{\rho} r^\alpha \right)  r \mathrm{d}{r} \right], \nonumber  \\
\end{IEEEeqnarray}
where ($b$) follows by the probability generating functional of $\Phi_e$\footnote{For a point process $\phi$, the probability generating functional is defined as $G_\phi(f) = \mathbb{E} \left[\prod _{x \in \phi }f(x) \right ]$ for $0<f(x) \leq 1$. If $\phi$ is a PPP with intensity function $\lambda(x)$, then $G_\phi(f)=\exp\left[-\int (1-f(x))\lambda(x))\mathrm{d}x\right ]$.}, ($c$) is based on Jensen's inequality, and ($d$) follows by converting Cartesian to Polar Coordinate and the tail distribution of $H_{e}$.

The lower bound of SOP is obtained by considering the success probability at the eavesdropper nearest to the transmitter. Denote the location of the nearest eavesdropper to the typical transmitter as $X_{e_1}$ and denote their distance as $r_e$, i.e., $r_e = |X_{e_1}|$. The probability density function of $r_e$ is given by
\begin{equation}\label{eq_PDF_nearestEve}
    f_{R_e}(r_e) = 2 \pi \lambda_e r_e \exp(-\pi \lambda_e r_e^2),
\end{equation}
which is the probability that no eavesdropper existing within the disk $\mathcal{B}(o,r_e)$ centered at $o$ with radius $r_e$.
The lower bound of SOP can be given by
\begin{IEEEeqnarray}{rCl}\label{eq_secOutPro_lb1_derivation}
\textrm{P}_\textrm{so}(\beta_e) &\geq& \mathbb{P}\left( \textrm{SINR}(X_{e_1}) \geq \beta_e \right ) \nonumber \\
&=& \int_{0}^{\infty}\mathbb{P} \left( \textrm{SINR}(X_{e_1}) \geq \beta_e \mid |X_{e_1}|= r_e\right ) f(r_e)\mathrm{d}r_e \nonumber \\
&=& \int_{0}^{\infty} \exp \left[- \theta \left(\frac{\beta_e}{\rho}\right)^{\frac{2}{\alpha}} r_e^2 \right] \mathcal{L}_{W} \left( \frac{\beta_e}{\rho} r_e^\alpha \right) f(r_e)\mathrm{d}r_e \nonumber \\
\end{IEEEeqnarray}
The lower bound in (\ref{eq_secOutPro_lb1}) follows by simplifying (\ref{eq_secOutPro_lb1_derivation}).
\end{IEEEproof}

From the definitions of COP and SOP, the following remark can be concluded.
\begin{remark}
The connection outage probability $\textrm{P}_\textrm{co} (\beta_t)$ increases with $\beta_t$, while the secrecy outage probability $\textrm{P}_\textrm{so} (\beta_e)$ decreases with $\beta_e$.
\end{remark}

Given the connection outage constraint $\sigma$, the codeword rate can be given by
\begin{equation}\label{eq_secRate}
    \mathcal{R}_t = \log \left( 1+ \textrm{P}_\textrm{co}^{-1}(\sigma) \right),
\end{equation}
where $\textrm{P}_\textrm{co}^{-1}$ is the inverse function of $\textrm{P}_\textrm{co}$ in (\ref{eq_conOutPro}).

Given the secrecy outage constraint $\epsilon$, the data rate cost against eavesdroppers can be given by
\begin{equation}\label{eq_eveRate}
    \mathcal{R}_e = \log \left( 1+ \textrm{P}_\textrm{so}^{-1}(\epsilon) \right) ,
\end{equation}
where $\textrm{P}_\textrm{so}^{-1}$ is the inverse function of $\textrm{P}_\textrm{so}$ in (\ref{eq_secOutPro_formula}).

The above inverse functions $\textrm{P}_\textrm{co}^{-1}$ and $\textrm{P}_\textrm{so}^{-1}$ exist because of the strict monotonicity of both connection and secrecy outage probabilities. For a given distribution of $W$, $\textrm{P}_\textrm{co}^{-1}$ can be numerically calculated based on (\ref{eq_conOutPro}),
and bounds of $\textrm{P}_\textrm{so}^{-1}$ can be numerically calculated based on the bounds in Lem. \ref{lemma_secOutPro}.

Based on the definition in Section \ref{section2_metric}, the STC can be derived in the following theorem.
\begin{theorem}\label{theorem_secTransCapacity}
The secrecy transmission capacity of the concerned wireless network with a connection outage constraint of $\sigma$ and a secrecy outage constraint of $\epsilon$ is given by
\begin{equation}\label{eq_secTransCapacity}
    \tau = (1-\sigma) \lambda_\textrm{T} \left[ \mathcal{R}_t - \mathcal{R}_e \right]^{+},
\end{equation}
where $\mathcal{R}_t$ and $\mathcal{R}_e$ are given in (\ref{eq_secRate}) and (\ref{eq_eveRate}). In particular, a lower bound of secrecy transmission capacity $\tau^{l}$  is derived when we use $\textrm{P}_\textrm{so}^{u}$ in (\ref{eq_secOutPro_ub1}) to calculate $\mathcal{R}_e$, while an upper bound of secrecy transmission capacity $\tau^{u}$ is derived when we use $\textrm{P}_\textrm{so}^{l}$ in (\ref{eq_secOutPro_lb1}) to calculate $\mathcal{R}_e$.
\end{theorem}
\begin{IEEEproof}
The STC can be directly derived by following the definition in Section \ref{section2_metric}. The potential problem is the existence of the inverse functions of $\textrm{P}_\textrm{so}^{u}$ and $\textrm{P}_\textrm{so}^{l}$. We now show that
$\textrm{P}_\textrm{so}^{u}$ has the unique inverse function; the existence of inverse function of $\textrm{P}_\textrm{so}^{l}$ can be proved in the similar way by showing that it is strictly monotonic. The derivative of $\textrm{P}_\textrm{so}^{u}$ is given by
\begin{IEEEeqnarray}{rCl}
   \IEEEeqnarraymulticol{3}{l}{
   \frac{\mathrm{d} \textrm{P}_\textrm{so}^{u} (\beta_e)}{\mathrm{d} \beta_e}
   } \nonumber \\
    &=& -  2 \pi \lambda_e  \exp\left( - 2 \pi \lambda_e \int_{0}^{\infty} e^{- \left( \frac{\beta_e}{\rho}\right )^{\frac{2}{\alpha}} \theta r^2 } \mathcal{L}_{W} \left(\frac{\beta_e}{\rho} r^\alpha \right) r \mathrm{d}{r} \right) \nonumber \\
   && \times \!\! \int_{0}^{\infty} \left [ \frac{2 \theta}{\alpha} \frac{{\beta_e}^{\frac{2}{\alpha} -1}} {\rho^\frac{2}{\alpha}} r^2 \mathcal{L}_{W} \left(\frac{\beta_e}{\rho} r^\alpha \right) \right. \nonumber \\
   && \left. + \frac{r^\alpha}{\rho} \int_{0}^{\infty} w e^{-\frac{\beta_e}{\rho} r^\alpha w}  f_W(w) \mathrm{d}{w} \right ] e^{- \left( \frac{\beta_e}{\rho}\right )^{\frac{2}{\alpha}} \theta r^2 } r \mathrm{d}{r}, \nonumber
\end{IEEEeqnarray}
where $f_W(w)$ is the probability density function of the random noise $W$.
The Laplace transform of $W$ is given by
\begin{equation}\label{eq_laplaceTran_noise}
    \mathcal{L}_{W} \left(\frac{\beta_e}{\rho} r^\alpha \right) = \int_{0}^{\infty} e^{-\frac{\beta_e}{\rho} r^\alpha w}  f_W(w) \mathrm{d}{w}.
\end{equation}
It is obvious that $\frac{\mathrm{d} \textrm{P}_\textrm{so}^{u} (\beta_e)}{\mathrm{d} \beta_e} <0$, which proves that $\textrm{P}_\textrm{so}^{u}$ has the unique inverse function.
\end{IEEEproof}

Notice that $\textrm{P}_\textrm{so}^{u}(\beta_e)$ derived in (\ref{eq_secOutPro_ub1}) will be shown to be very tight by simulation (see Figs. \ref{fig:secOutPro_Noise} and \ref{fig:secOutPro_TxDen}),
and that other results derived based on the same bounding techniques have also been illustrated to be tight in \cite{Ganti2007, Zhou2011b}.
Moreover, the lower bound of secrecy transmission capacity $\tau^{l}$ derived based on $\textrm{P}_\textrm{so}^{u}(\beta_e)$ gives a very tight approximation of the exact value of $\tau$.

\subsection{$W = w$ \& $\alpha=4$}
We now consider the special scenario when the path-loss exponent $\alpha=4$ and noise power $W$ is a fixed value $w$ across space and time slot, i.e., $W = w$, and derive the closed-form expressions.

When the noise is a constant $w$, we can derive the COP and SOP by replacing $\mathcal{L}_{W} \left( \frac{\beta_t}{\rho} L^\alpha \right)$ by $\exp \left[ - \frac{\beta_t}{\rho} w L^\alpha \right]$ into (\ref{eq_conOutPro}) and Lem. \ref{lemma_secOutPro}.

When the noise power is a constant $w$ for each time slot and $\alpha=4$, the COP is given by
\begin{equation} \label{eq_conOutPro_alpha4}
    \textrm{P}_\textrm{co}(\beta_t) =  1- \exp \left[ -  w \frac{\beta_t}{\rho}L^4 - \theta \left(\frac{\beta_t}{\rho}\right)^{\frac{1}{2}} L^2 \right].
\end{equation}
Therefore, for a connection outage constraint $\textrm{P}_\textrm{co}(\beta_t) = \sigma$, we have
\begin{equation} \label{eq_conSINRthresh_alpha4}
    \beta_t =  \textrm{P}_\textrm{co}^{-1}(\sigma) = \rho \left( \frac{-\theta + \sqrt{\theta^2 + 4 w \ln \frac{1}{1-\sigma}}}{2 w L^2} \right)^2.
\end{equation}

\begin{corollary}
For the constant noise $w$ and $\alpha=4$, the tight upper bound and lower bound of the secrecy outage probability are given by
\begin{equation} \label{eq_secOutPro_ub1_alpha4}
    \textrm{P}_\textrm{so}^{u}(\beta_e) = 1 - \exp \!\! \left[ - \frac{\pi^{\frac{3}{2}}\lambda_e}{2} \sqrt{\frac{\rho}{\beta_e w}} \exp \!\! \left(\! \frac{\theta^2}{4 w} \! \right) \mathrm{Erfc} \! \left( \! \frac{\theta}{2 \sqrt{w}} \! \right ) \! \right]
\end{equation}
and
\begin{IEEEeqnarray}{rCl} \label{eq_secOutPro_lb1_alpha4}
    \textrm{P}_\textrm{so}^{l}(\beta_e) &=& \frac{\pi^{\frac{3}{2}}\lambda_e}{2}  \sqrt{\frac{\rho}{\beta_e w}}
\exp \!\! \left[ \frac{ \left( \theta \sqrt{\frac{\beta_e}{\rho}}+ \pi \lambda_e \right) ^2}{4 w \frac{\beta_e}{\rho}} \right] \nonumber \\
&& \mathrm{Erfc} \! \left( \frac{\theta \sqrt{\frac{\beta_e}{\rho}}+ \pi \lambda_e}{2 \sqrt{w \frac{\beta_e}{\rho}}} \right ),
\end{IEEEeqnarray}
where $\mathrm{Erfc}(z)=\frac{2}{\sqrt{\pi}}\int_{z}^{\infty}e^{-t^2}\mathrm{d}{t}$ is the complementary error function.
\end{corollary}
\begin{IEEEproof}
Replacing $\mathcal{L}_{W} \left( \frac{\beta_t}{\rho} L^\alpha \right)$ by $\exp \left[ - \frac{\beta_t}{\rho} w L^\alpha \right]$ and $\alpha=4$ into (\ref{eq_secOutPro_ub1}) and (\ref{eq_secOutPro_lb1}), we can derive the above results based on the following identity
\begin{equation}\label{identity1}
    \int_{0}^{\infty}e^{-a t^4 -b t^2} t \mathrm{d}{t} = \frac{\sqrt{\pi}}{4\sqrt{a}} \exp \left( \frac{b^2}{4a}\right) \mathrm{Erfc} \left(\frac{b}{2\sqrt{a}}\right).
\end{equation}
\end{IEEEproof}

For a secrecy outage constraint $\textrm{P}_\textrm{so}^{u}(\beta_e) = \epsilon$, we have
\begin{equation} \label{eq_secSINRthresh_alpha4}
    \beta_e = \frac{\rho}{w} \left[ \frac{ \pi^{\frac{3}{2}} \lambda_e \exp \!\! \left(\! \frac{\theta^2}{4w} \! \right) \mathrm{Erfc} \! \left( \! \frac{\theta}{2 \sqrt{w}} \! \right )}{2 \ln \frac{1}{1 - \epsilon}} \right]^2,
\end{equation}
which is an upper bound of the eavesdropper's decoding threshold under the secrecy constraint of $\epsilon$.

\begin{theorem}\label{theorem_secTransCapacity_alpha4}
For the constant noise $w$ and $\alpha=4$, the tight lower bound of secrecy transmission capacity $\tau^u$ with a connection outage constraint of $\sigma$ and a secrecy outage constraint of $\epsilon$ is given by
\begin{equation}\label{eq_secTransCapacity_alpha4}
    \tau^u = (1-\sigma) \lambda_\textrm{T} \left[ \log \left(1 + \beta_t \right) - \log \left(1 + \beta_e \right) \right]^{+},
\end{equation}
where $\beta_t$ and $\beta_e$ are given in (\ref{eq_conSINRthresh_alpha4}) and (\ref{eq_secSINRthresh_alpha4}).
\end{theorem}

\begin{corollary}
For the constant noise $w$ and $\alpha=4$, the condition for a positive secrecy transmission capacity is given by
\begin{equation}\label{eq_positiveCapacityCondi_alpha4}
    \frac{\left( -\theta + \sqrt{\theta^2 + 4 w \ln \frac{1}{1-\sigma}} \right) \ln \frac{1}{1-\epsilon}}{ \pi^{\frac{3}{2}} r^2 \lambda_e \sqrt{w} \exp \!\! \left(\! \frac{\theta^2}{4 w} \! \right) \mathrm{Erfc} \! \left( \! \frac{\theta}{2 \sqrt{w}} \! \right )} > 1.
\end{equation}
\end{corollary}

\section{Second Model: Random Transmission Distance} \label{section4}
In this section, we consider a more realistic transmission scenario of random transmission distance.
In particular, we consider the nearest-receiver transmission (NRT) scheme.

Recall $\Phi_\textrm{T}$ is the PPP of intensity $\lambda_\textrm{T}$ of transmitters, and $\Phi_\textrm{R}$ is the PPP of intensity $\lambda_\textrm{R}$ of potential receivers. We now consider the case that each transmitter adopts NRT, i.e., each transmitter transmits to its nearest receiver. For simplicity, we ignore the failures caused by the fact that multiple transmitters may select the same receiver \cite{Weber2007, Bacceli2010}.

Denoting the distance from the typical transmitter to its nearest receiver in $\Phi_\textrm{R}$ by $R$, the probability density function of $R$ is given by
\begin{equation}\label{eq_PDF_NRT}
    f_R(r) = 2 \pi \lambda_\textrm{R} r \exp(-\pi \lambda_\textrm{R} r^2).
\end{equation}

It is noticed that the transmit distance impacts the TOP, while it has no impact on the SOP.
\begin{lemma} \label{lemma_conOutPro_NRT}
For the concerned wireless network with network parameters $\lambda_\textrm{T}$, $\lambda_\textrm{R}$, $W$ and $\alpha$, and transmission power $\rho$ defined above,
its connection outage probability under NRT
for a given receiver's SINR threshold $\beta_t$
is determined by
\begin{equation} \label{eq_conOutPro_NRT}
    \textrm{P}_{\textrm{co},\textrm{n}} (\beta_t) = 1- 2 \pi \lambda_\textrm{R} \int_{0}^{\infty} \!\! e^ {- \left[ \theta \left(\frac{\beta_t}{\rho}\right)^{\frac{2}{\alpha}} +\pi \lambda_\textrm{R} \right] r^2} \mathcal{L}_{W} \left( \frac{\beta_t}{\rho} r^\alpha \right) r \mathrm{d}{r},
\end{equation}
where $ \theta = \pi \lambda_\textrm{T} \Gamma( 1- 2/\alpha ) \Gamma( 1+ 2/\alpha )$ is given in Lem. \ref{lemma_secOutPro}.
\end{lemma}
\begin{IEEEproof}
$\textrm{P}_{\textrm{co},\textrm{n}} (\beta_t)$ can be derived based on the following formula,
\begin{equation}\label{eq_conOutPro_NRT_derivation}
    \textrm{P}_{\textrm{co},\textrm{n}}(\beta_t)
= \int_{0}^{\infty} \textrm{P}_\textrm{co} (\beta_t) f_R(r) \mathrm{d}r,
\end{equation}
where $\textrm{P}_\textrm{co} (\beta_t)$ is the COP for a fixed transmit distance derived in Section \ref{section3}.
\end{IEEEproof}

The STC for nearest receiver transmission can be given as follows.
\begin{theorem}\label{theorem_secTransCapacity_NRT}
The secrecy transmission capacity under the nearest receiver transmission (NRT) with a connection outage constraint of $\sigma$ and a secrecy outage constraint of $\epsilon$ is given by
\begin{equation}\label{eq_secTransCapacity_NRT}
    \tau_\textrm{n} = (1-\sigma) \lambda_\textrm{T} \left[ \mathcal{R}_t - \mathcal{R}_e \right]^{+},
\end{equation}
where $\mathcal{R}_t = \log \left(1 + \textrm{P}_{\textrm{co},\textrm{n}}^{-1} (\sigma) \right)$ and $\mathcal{R}_e$ is given in (\ref{eq_eveRate}).
In particular, a lower bound of secrecy transmission capacity $\tau^{l}_\textrm{n}$  is derived when we use $\textrm{P}_\textrm{so}^{u}$ in (\ref{eq_secOutPro_ub1}) to calculate $\mathcal{R}_e$, while an upper bound of secrecy transmission capacity $\tau^{u}_\textrm{n}$ is derived when we use $\textrm{P}_\textrm{so}^{l}$ in (\ref{eq_secOutPro_lb1}) to calculate $\mathcal{R}_e$.
\end{theorem}
\begin{IEEEproof}
The STC can be directly derived by following the definition in Section \ref{section2_metric}. The potential problem is the existence of the inverse function $\textrm{P}_{\textrm{co},\textrm{n}}^{-1}$.  The derivative of $\textrm{P}_{\textrm{co},\textrm{n}}$ is given by
\begin{IEEEeqnarray}{rCl}
    \IEEEeqnarraymulticol{3}{l}{
   \frac{\mathrm{d} \textrm{P}_{\textrm{co},\textrm{n}} (\beta_t)}{\mathrm{d} \beta_t}
   } \nonumber \\
    &=& 2 \pi \lambda_\textrm{R} \int_{0}^{\infty} e^{- \left[ \theta \left(\frac{\beta_t}{\rho}\right)^{\frac{2}{\alpha}} +\pi \lambda_\textrm{R} \right] r^2 } \left [  \frac{2 \theta}{\alpha} \frac{{\beta_t}^{\frac{2}{\alpha} -1}} {\rho^\frac{2}{\alpha}} r^2 \mathcal{L}_{W} \left(\frac{\beta_t}{\rho} r^\alpha \right) \right.
    \nonumber \\
    && \left. + \vphantom{\frac{{\beta_t}^{\frac{2}{\alpha} -1}} {\rho^\frac{2}{\alpha}}}  \: \frac{r^\alpha}{\rho} \int_{0}^{\infty} w e^{-\frac{\beta_t}{\rho} r^\alpha w}  f_W(w) \mathrm{d}{w}
     \right ]  r \mathrm{d}{r},
\end{IEEEeqnarray}
where $f_W(w)$ is the probability density function of the random noise $W$.
It is obvious that $\frac{\mathrm{d} \textrm{P}_{\textrm{co},\textrm{n}} (\beta_t)}{\mathrm{d} \beta_t} >0$, which proves that $\textrm{P}_{\textrm{co},\textrm{n}}$ has the unique inverse function $\textrm{P}_{\textrm{co},\textrm{n}}^{-1}$.
\end{IEEEproof}

\section{Numeric Analysis and Discussion} \label{section5}
In this section, we first verify the efficiency of the theoretical models of connection outage probability and secrecy outage probability through simulation, and then explore the inherent tradeoffs among different system parameters.

\subsection{Model Validation} 
We developed a simulator, which is now available at \cite{Zhu2013a}, to simulate the message transmission process under the system model defined in Section \ref{section_systemModel} and the transmission schemes defined at the beginnings of Section \ref{section3} and Section \ref{section4} for fixed and random transmission distances, respectively.
To model the large-scale network, the network size was set to $100 \times 100$ for $\lambda_\textrm{T} \geq 10^{-3} $, and $300 \times 300$ for $10^{-4} \leq \lambda_\textrm{T} < 10^{-3} $ \cite{Haenggi2009a}.
The performance of the network is considered on an additional transmitter located at the center of the network.
Specifically, we considered the COP and SOP of the typical transmitter.
It is noticed that the efficiency of the STC relies on the efficiencies of the COP and SOP.

\begin{figure}[tb]
\centering
\includegraphics[width= 0.9 \linewidth]{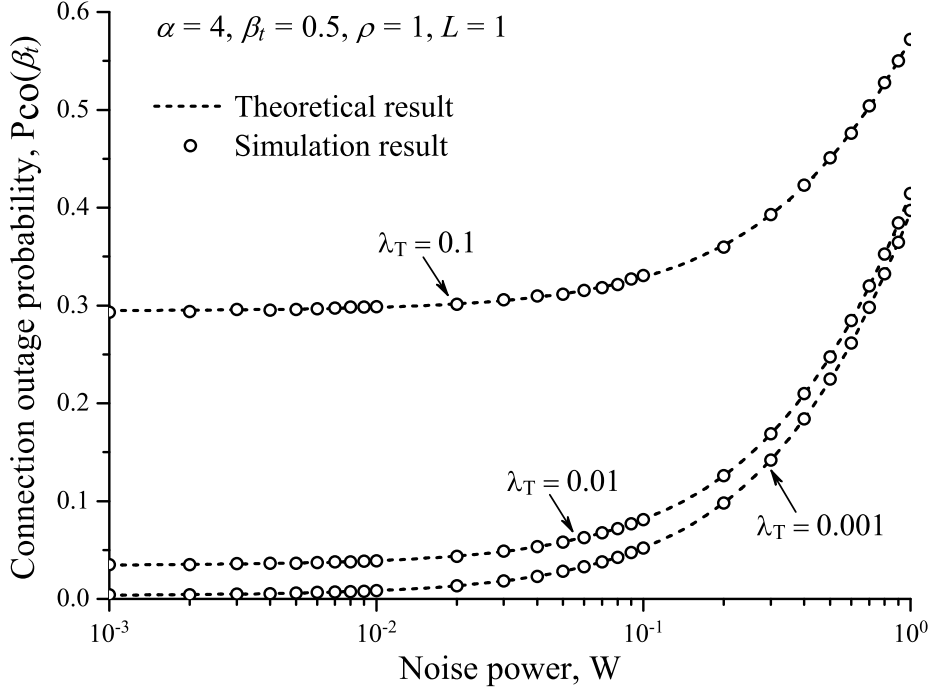}
\caption{Connection outage probability $\textrm{P}_\textrm{co}$ vs. noise power $W$ when $r = 1$.}
\label{fig:conOutPro_Noise}
\end{figure}

\begin{figure}[tb]
\centering
\includegraphics[width= 0.9 \linewidth]{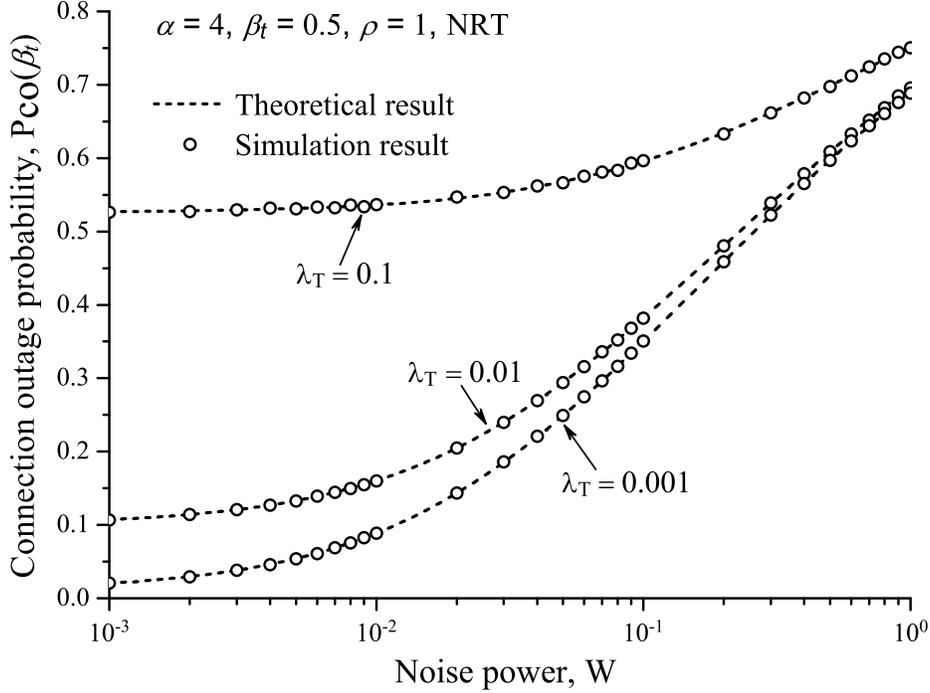}
\caption{Connection outage probability $\textrm{P}_\textrm{co}$ vs. noise power $W$ when NRT.}
\label{fig:conOutPro_Noise_NRT}
\end{figure}

To validate the COP, we considered networks with $\alpha = 4$, $\rho=1$, $\beta_t = 0.5$ and several different settings of transmitter density (i.e., $\lambda_\textrm{T} = \{ 0.1, 0.01, 0.001 \}$) in Figs. \ref{fig:conOutPro_Noise} and \ref{fig:conOutPro_Noise_NRT}.
In particular, Fig. \ref{fig:conOutPro_Noise} validates the COP for the fixed transmission distance of $L=1$, and Fig. \ref{fig:conOutPro_Noise_NRT} validates the COP for the nearest neighbor transmission.
It can be observed from Figs. \ref{fig:conOutPro_Noise} and \ref{fig:conOutPro_Noise_NRT} the simulation results match the theoretical ones very well, which validates the efficiencies of our theoretical models of COP for both fixed and random transmission distances.

To validate the SOP, we considered networks with $\alpha = 4$, $\rho=1$, $\beta_e = 0.1$ and eavesdropper density $\lambda_e = 0.001$ in Figs. \ref{fig:secOutPro_Noise} and \ref{fig:secOutPro_TxDen}.
In particular, Fig. \ref{fig:secOutPro_Noise} validates the SOP for a transmitter density of $\lambda_\textrm{T} = 0.01$ and different settings of noise power,
and Fig. \ref{fig:secOutPro_TxDen} validates the SOP for a noise power of $W=0.001$ and different settings of transmitter density.
The results in Figs. \ref{fig:secOutPro_Noise} and \ref{fig:secOutPro_TxDen} indicate that the upper and lower bounds of SOP derived in this paper are tight, and that the upper bound is very close to the simulated SOP.
In Figs. \ref{fig:secOutPro_Noise} and \ref{fig:secOutPro_TxDen}, the solid lines show the previous upper bound of SOP for interference-limited network in \cite{Zhou2011b}. It is obvious that the previous upper bound is very loose for network scenarios where noise cannot be  neglected.

\subsection{Outage Performances vs. Noise and Interference}
We now explore the impacts of noise and interference on the COP.
We can see from Figs. \ref{fig:conOutPro_Noise} and \ref{fig:conOutPro_Noise_NRT} that the connection outage probability $\textrm{P}_\textrm{co}$ increases with the noise power $W$, which indicates that noise deteriorates the reliability performance in the sense that the legitimate receiver can recover messages successfully.
For a given $W$, we can also observe from Figs. \ref{fig:conOutPro_Noise} and \ref{fig:conOutPro_Noise_NRT} that $\textrm{P}_\textrm{co}$ becomes greater for a larger transmitter density $\lambda_\textrm{T}$.
This indicates that also interference deteriorates the reliability performance.

\begin{figure}
\centering
\includegraphics[width= 0.9 \linewidth]{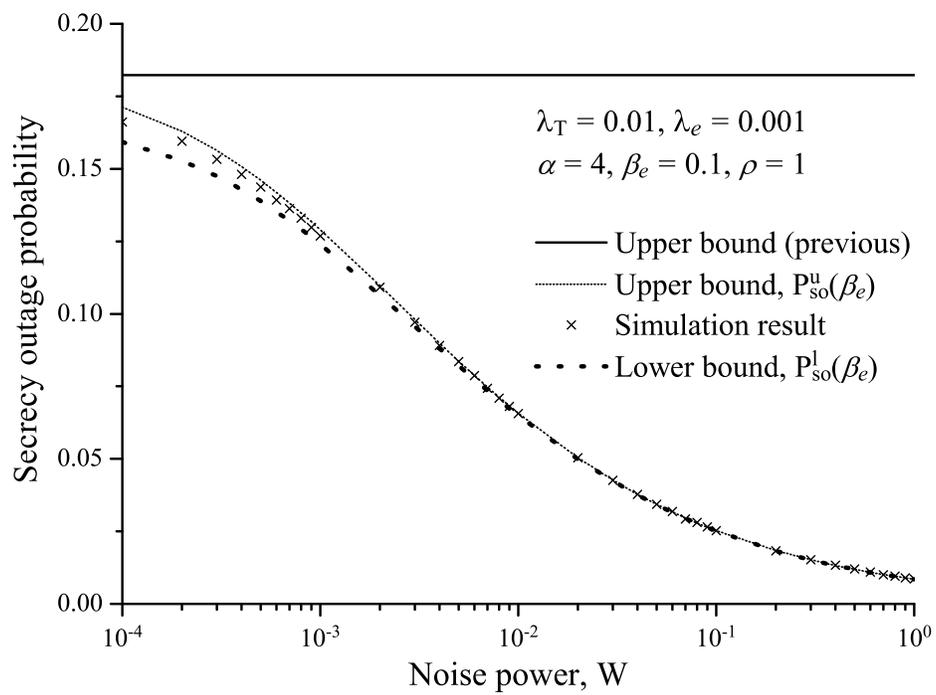}
\caption{Connection outage probability $\textrm{P}_\textrm{co}$ vs. noise power $W$ when $r = 1$.}
\label{fig:secOutPro_Noise}
\end{figure}

\begin{figure}
\centering
\includegraphics[width= 0.9 \linewidth]{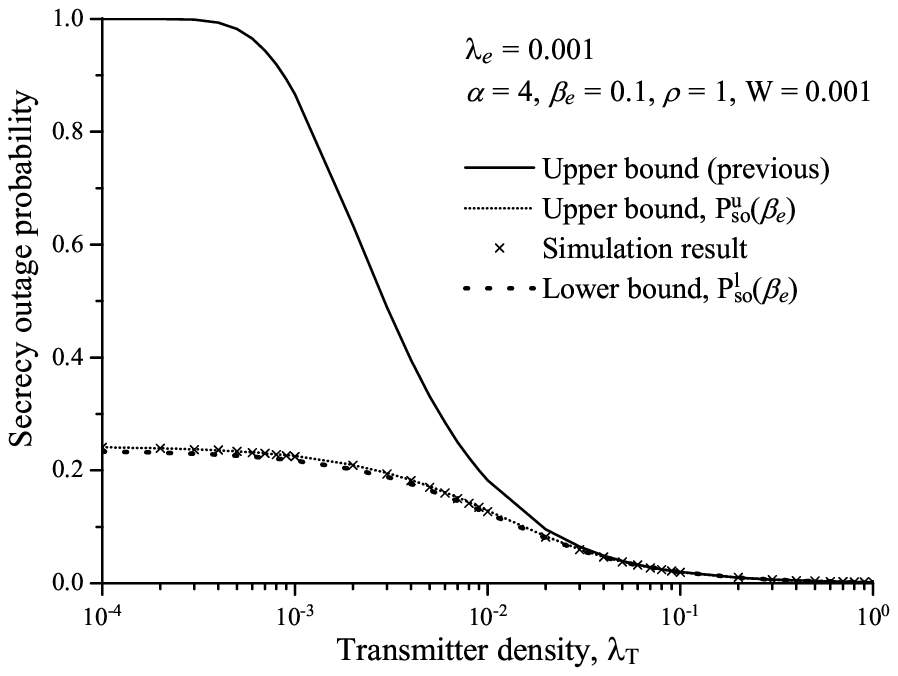}
\caption{Connection outage probability $\textrm{P}_\textrm{co}$ vs. noise power $W$ when $r = 1$.}
\label{fig:secOutPro_TxDen}
\end{figure}

To illustrate the impacts of noise and interference on the SOP, we summarize in Fig. \ref{fig:secOutPro_Noise} how the secrecy outage probability $\textrm{P}_\textrm{so}$ varies with $W$, and summarize in Fig. \ref{fig:secOutPro_TxDen} how $\textrm{P}_\textrm{so}$ varies with $\lambda_\textrm{T}$.
We can see from Figs. \ref{fig:secOutPro_Noise} and \ref{fig:secOutPro_TxDen} that $\textrm{P}_\textrm{so}$ decreases with either $W$ or $\lambda_\textrm{T}$,
which indicates that both noise and interference help the security performance in the sense that eavesdroppers cannot recover messages successfully.

\subsection{Secrecy Transmission Capacity vs. Noise and Interference}
\begin{figure}[!h]
\centering
\includegraphics[width= 0.9 \linewidth]{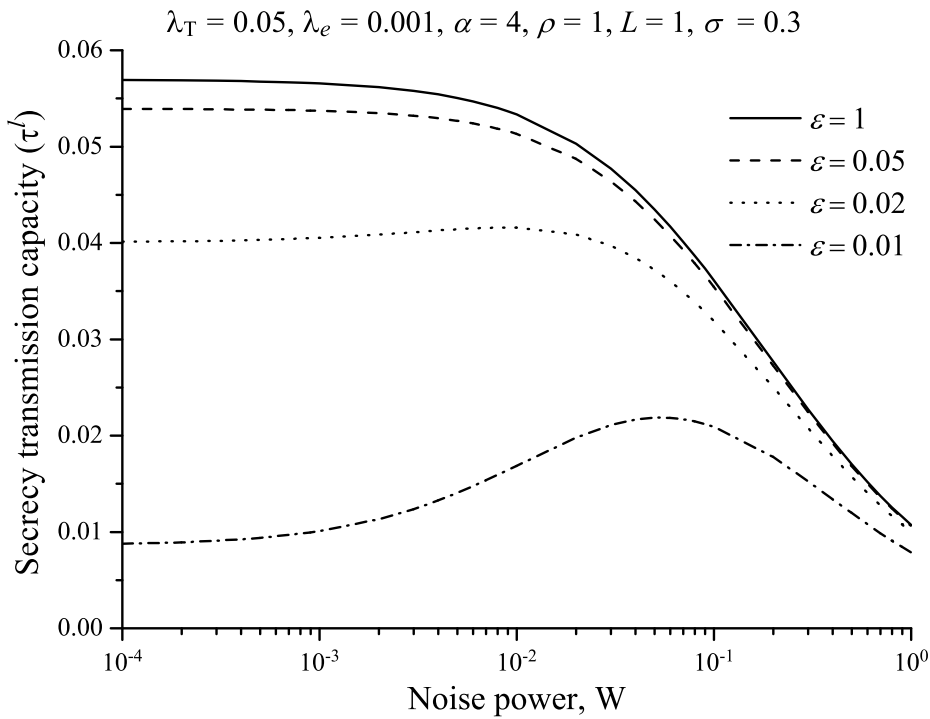}
\caption{Connection outage probability $\textrm{P}_\textrm{co}$ vs. noise power $W$ when $r = 1$.}
\label{fig:capacity_Noise}
\end{figure}

\begin{figure}
\centering
\includegraphics[width= 0.9 \linewidth]{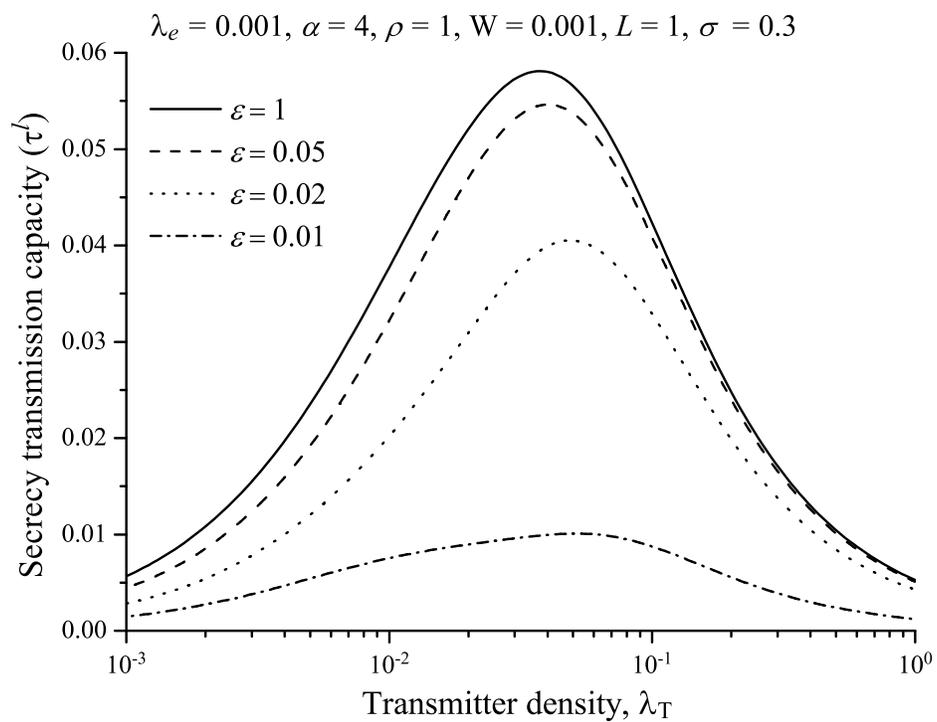}
\caption{Connection outage probability $\textrm{P}_\textrm{co}$ vs. noise power $W$ when $r = 1$.}
\label{fig:capacity_TxDen}
\end{figure}

To further explore the impacts of noise and interference on the STC, we show in Fig. \ref{fig:capacity_Noise} how the (lower bound of) secrecy transmission capacity $\tau^{l}$ varies with $W$,
and summarize in Fig. \ref{fig:capacity_TxDen} how $\tau^{l}$ varies with $\lambda_\textrm{T}$.
Notice that, although the lower bound of secrecy transmission capacity $\tau^{l}$ is adopted here, we can get the same conclusions about the impacts of  noise and interference on the exact secrecy transmission capacity,
since $\tau^{l}$ is very close to $\tau$.
It can be observed from Fig. \ref{fig:capacity_Noise} that $\tau^{l}$ first increases with $W$ and then decreases with $W$.
It is noticed that the overall impact of noise on $\tau^{l}$ composes both impacts of noise on $\textrm{P}_\textrm{co}$ and $\textrm{P}_\textrm{so}$.
The above phenomenon is due to that the helpful impact of noise on $\textrm{P}_\textrm{so}$ dominates the overall impact of noise on $\tau^{l}$ at first,
and the harmful impact of noise on $\textrm{P}_\textrm{co}$ dominates the overall impact of noise on $\tau^{l}$ after the optimum $W$.
Therefore, it is suggested to add some artificial noise to achieve a larger STC for some occasions.
From Fig. \ref{fig:capacity_TxDen}, we also find that $\tau^{l}$ first increases with $\lambda_\textrm{T}$ and then decreases with $\lambda_\textrm{T}$.
The reason for such a phenomenon is similar as the one for the impact of noise.


\subsection{Impacts of Secrecy on Secrecy Transmission Capacity}
\begin{figure}[tb]
\centering
\includegraphics[width= 0.9 \linewidth]{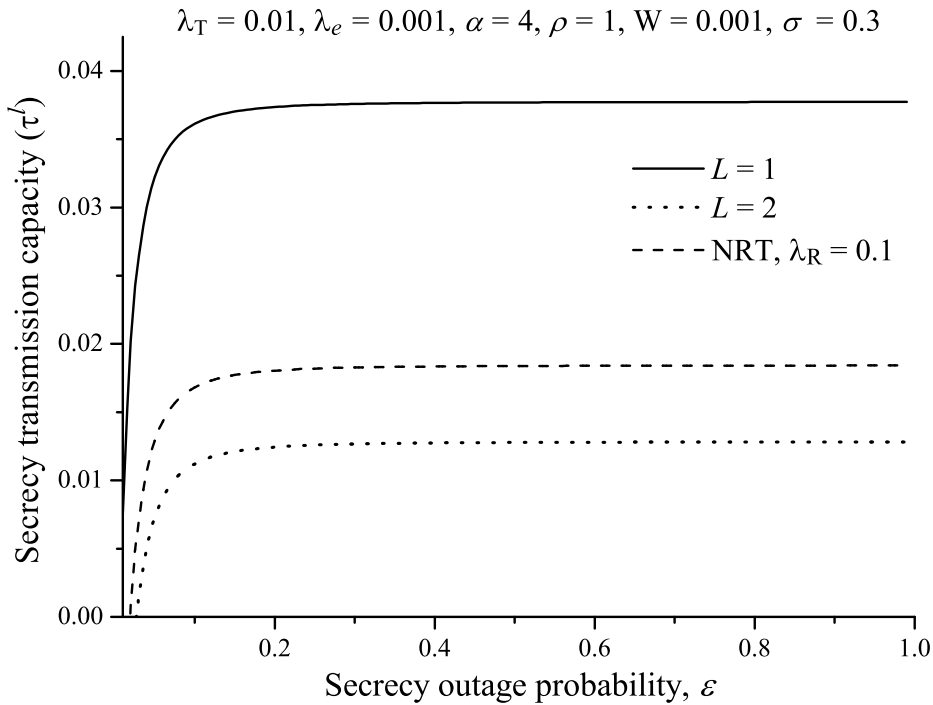}
\caption{Connection outage probability $\textrm{P}_\textrm{co}$ vs. noise power $W$ when $r = 1$.}
\label{fig:capacity_secOutPro}
\end{figure}

To understand the impact of secrecy on STC, we show in Fig. \ref{fig:capacity_secOutPro} how $\tau^{l}$ varies with the secrecy outage constraint $\textrm{P}_\textrm{so} (\beta_e) = \epsilon$ for the scenarios of $\lambda_\textrm{T}=0.01$, $\lambda_e = 0.001$, $\alpha = 4$, $\rho=1$, $\beta_e = 0.1$ and different transmission schemes (i.e., fixed distance of $L = {1, 2}$ or NRT of $\lambda_\textrm{R}=0.1$).
Fig. \ref{fig:capacity_secOutPro} shows that $\tau^{l}$ increases with $\epsilon$ sharply when $\epsilon$ is small while increases with $\epsilon$ slowly when $\epsilon$ is large.
For example, there is an over $85\%$ increment in $\tau^{l}$ by relaxing the secrecy constraint from $\epsilon =0.02$ to  $\epsilon =0.1$ for NRT with $\lambda_\textrm{R}=0.1$, but only less than $15\%$ increment from $\epsilon =0.1$ to $\epsilon =1$.
This indicates that the performance of STC can be improved a lot by allowing small probability of secrecy outage.
Furthermore, it is noticed that the impact of secrecy on STC is the same for fixed or random distance transmissions.

\section{Conclusion} \label{section6}
This paper studied the secrecy transmission capacity in noisy wireless ad hoc networks, where both background noise and interference from concurrent transmitters affect the received signals, which cover the previous result of secrecy transmission capacity in interference-limited networks as a special case \cite{Zhou2011b}.
Based on the tools from stochastic geometry, we first focused on a basic scenario where the transmission distances are assumed to be the same for all the transmitters, and derived the exact connection outage probability, and bounds of secrecy outage probability and secrecy transmission capacity.
We then extend our analysis to a more realistic transmission scenario where each transmitter transmits to its nearest receiver.
The simulation has also been conducted to verify the efficiency of our theoretical ones.
It is notable that the upper bound of secrecy outage probability or lower bound of secrecy transmission capacity has been shown very tight.
Another interesting finding is that a proper amount of noise is helpful to the secrecy transmission capacity while too much noise is harmful.
Therefore, to achieve a larger secrecy transmission capacity, it is sometimes good to add some artificial noise.





\bibliographystyle{model1-num-names}
\bibliography{ref}







\end{document}